# Challenges in first-principles *NPT* molecular dynamics of Soft Porous Crystals: a case study on MIL-53(Ga)

Volker Haigis[a)] and Yacine Belkhodja
*CNRS-ENS-UPMC, Département de Chimie, École Normale Supérieure, 24 rue Lhomond, 75005 Paris, France.*

François-Xavier Coudert
*Institut de Recherche de Chimie Paris, CNRS – Chimie ParisTech, 11 rue Pierre et Marie Curie, 75005 Paris, France.*

Rodolphe Vuilleumier and Anne Boutin
*CNRS-ENS-UPMC, Département de Chimie, École Normale Supérieure, 24 rue Lhomond, 75005 Paris, France.*

Soft porous crystals present a challenge to molecular dynamics simulations with flexible size and shape of the simulation cell (i.e., in the $NPT$ ensemble), since their framework responds very sensitively to small external stimuli. Hence, all interactions have to be described very accurately in order to obtain correct equilibrium structures. Here, we report a methodological study on the nanoporous metal-organic framework MIL-53(Ga), which undergoes a large-amplitude transition between a narrow- and a large-pore phase upon a change in temperature. Since this system has not been investigated by density functional theory (DFT)-based $NPT$ simulations so far, we carefully check the convergence of the stress tensor with respect to computational parameters. Furthermore, we demonstrate the importance of dispersion interactions and test two different ways of incorporating them into the DFT framework. As a result, we propose two computational schemes which describe accurately the narrow- and the large-pore phase of the material, respectively. These schemes can be used in future work on the delicate interplay between adsorption in the nanopores and structural flexibility of the host material.

## I. INTRODUCTION

Soft Porous Crystals are a fascinating subclass of metal–organic frameworks which behave in a remarkable stimuli-responsive fashion.[1,2] Like all Metal–Organic Frameworks (MOFs), they are crystalline microporous materials whose three-dimensional framework is constructed from metal centers linked together by organic ligands, and thus present a large structural diversity and chemical versatility, enabling the design of new materials with tunable host–guest properties. Moreover, Soft Porous Crystals (SPCs) display reversible single-crystal-to-single-crystal structural transformations of large amplitude under a number of external physical constraints such as guest adsorption, temperature or mechanical pressure.[3,4] The number of such materials reported in the literature is rapidly growing, and they have potential applications in nanobiotechnology,[5] sensing for detecting traces of organic molecules,[2] slow release of drugs for long-release single-injection therapies,[6] and specific gas separations.[7,8]

In the last decade, a large range of theoretical chemistry techniques have been used with success to understand (and sometimes predict) the behavior of MOFs in general, and SPCs in particular, as well as their response to adsorption of guest molecules, changes in temperature or mechanical stress. The methods used in the literature to address these questions can be grouped in four different classes. The first one is the use of macroscopic thermodynamic models, using input from both experimental data and other theoretical calculations (for a recent review of these efforts, see Coudert et al.[4]). The second class of methods is that of ''static'' quantum chemistry calculations, giving insight at the microscopic scale on host properties and host–guest interactions. Quantum-chemical energy calculations and energy minimizations have been heavily used to help determination of experimental structures,[9] to shed light onto the energetics of host–guest interactions (e.g., adsorption enthalpies),[10,11] as well as those of structural changes[12,13] and elastic[14,15] properties of the host phase itself. However, while they can yield quantitative *ab initio* predictions of energies, such ''zero Kelvin'' methods fail to describe the finite-temperature dynamics and entropic effects that can play a crucial role in structural transitions in Soft Porous Crystals.[12,16]

A third class of methods used in the existing literature on adsorption-induced deformation of SPCs is that of forcefield-based molecular simulations methods, and in particular the classical Molecular Dynamics (MD) and Monte Carlo (MC) techniques. These simulations fully explore the phase space of the system under study (or aim at fully exploring it), thus providing full statistical mechanical information including entropic effects. MD simulations also enables one to access dynamical properties of both the host material (linker orientation dynamics, structural transitions, etc.) and the adsorbed phase (orientational dynamics, diffusion and transport properties). Moreover, MD and MC simulations can be performed in a variety of thermodynamic ensembles, mimicking different experimental conditions: $(N, V, T)$ for constant volume systems, $(N, \sigma, T)$ (where $\sigma$ is the stress tensor) for isobaric or iso-stress conditions in the absence of guest (or at fixed loading),[17] and the osmotic ensemble $(N_{\text{host}}, \mu_{\text{ads}}, \sigma, T)$ for adsorption-induced structural transitions.[18] The downside of these forcefield-based methods is that they rely on an empirical approximations of both the intramolecular and intermolecular interactions in the system. The design of these forcefields is a difficult and time-

---

[a)]Electronic mail: volker.haigis@ens.fr

consuming task, and is especially daunting in the case of flexible molecular materials, owing to the complexity of the intramolecular interactions that are to be modeled. As a consequence, while such methods have been used with success to describe structural transitions in SPCs such as MIL-53(Cr)[17] and DMOF-1,[19] they have only been applied to a small number of MOF materials or {MOF, guest} couples.[20,21]

The fourth class of methods is that of First-Principles Molecular Dynamics (FPMD), also called *ab initio* MD. In this approach, one performs finite-temperature MD using interatomic forces calculated through first-principles electronic structure methods. This approach is more generic than forcefield-based MD, since unbiased forces are obtained without the need of parameterizing a forcefield for each new system studied, or each new structure of a material. However, it has a higher computational cost, meaning that the time and length scales reached are more limited than those of classical MD. Nevertheless, recent advances in high-performance computing resources and parallelization of these software have enabled their use on complex molecular materials, including studies of adsorbed phases in microporous systems. Examples include zeolites[22] as well as a number of different MOFs.[23–26] FPMD also allows one to model certain electronic properties (e.g., dipole moment of adsorbed molecules[27]) that are hardly accessible through forcefield-based MD, as well as study chemical reactions, as was recently demonstrated in studies of thermal stability of materials from the IRMOF family.[28,29]

FPMD is naturally well-suited to study stimuli-induced deformations and structural transitions in flexible materials, especially when used in the isobaric ensemble to study pressure- and temperature-induced transitions. Yet, in spite of its recent successes in the field of MOFs, constant-pressure FPMD has been little used so far in the study of SPC.[25,26] In this paper, we describe why the use of first-principles molecular dynamics for flexible MOFs is a still a challenge today. We highlight both the theoretical and practical pitfalls of using the method on a tricky test case: the bistable MIL-53(Ga) "breathing" MOF. We show how crucial it is to carefully choose basis set, plane-wave cutoff, exchange–correlation functional and to account for dispersive interactions in order for FPMD simulations to describe the known experimental behavior properly. However, we stress that we do not aim at mapping the thermodynamic phase diagram, including the breathing behavior, of this material completely from first principles. Breathing transitions are cooperative, rare events, and reaching thermodynamic equilibrium in $NPT$ simulations of MOFs probably requires simulation times of at least several hundreds of picoseconds[30,31], which is currently not feasible with *ab initio* methods. Our objective is rather to establish a methodology for studying the narrow- and open-pore phases of MIL-53(Ga) at time scales accessible to *ab initio* MD, i.e., at most several tens of picoseconds. This requires to identify computational settings which keep the experimentally observed structures stable over this time scale. Once such a methodology is found, we intend to apply it to the study of the dynamics of gas adsorption and of the hydrothermal degradation of MOFs.

## II. MIL-53(GA) AS A TEST CASE FOR SIMULATIONS

The Soft Porous Crystal MIL-53(Ga) has the chemical formula $Ga(OH)(O_2C-C_6H_4-CO_2)$, with four formula units per unit cell, and consists of metal hydroxide chains –Ga–OH–Ga–OH– connected to each other by benzenedicarboxylate linkers[32]. It thus forms a three-dimensional framework with diamond-shaped channels running parallel to the inorganic chains and which can accomodate adsorbate molecules (see Fig. 4, upper panel, for a snapshot of the structure). The empty material (i.e. without adsorbate) takes on a monoclinic narrow-pore form with space group $C2/c$ at room temperature and transforms to an orthorhombic large-pore phase with space group *Imma* above 500 K[33]. The topology of the compound does not change, however, on this transformation. When exposed to ambient air at room temperature, the empty material transforms to a hydrated narrow-pore phase with one adsorbed $H_2O$ molecule per Ga and with channels slightly more open than in the empty narrow-pore phase. The lattice parameter $b$, representing one of the two pore diagonals, can be considered a phase indicator: it changes considerably upon the transition, from approximately 7 Å in the narrow-pore to more than 13 Å in the large-pore phase.

These structural transitions result from a delicate interplay of different interactions: while dispersion forces favor short distances between the organic linkers, i.e. narrow pores, coordination chemistry of octahedrally coordinated Ga favors the open structure[13]. Moreover, the adsorption of guest molecules can induce a shrinkage or an expansion of the material, depending on the size difference between the empty pores and the adsorbate[34]. Since Soft Porous Crystals are characterized by their particular sensitivity to external stimuli such as temperature changes and guest adsorption, the different factors determining the structure of MIL-53(Ga) have to be described accurately in order to reproduce the experimentally observed behavior. MIL-53(Ga) therefore represents a challenging test case for FPMD simulations in the $NPT$ ensemble, i.e. at fixed temperature and with flexible size and shape of the simulation cell.

## III. PLANE-WAVE CUTOFF AND GAUSSIAN BASIS SETS

All simulations were carried out with the CP2K package[35] in the framework of density functional theory (DFT) as implemented in the QUICKSTEP module[36]. This computer code uses atom-centered Gaussian basis sets to describe the Kohn-Sham orbitals, whereas the electronic density is represented in an auxiliary plane-wave basis. Hence, the convergence of the atomic forces and of the stress tensor has to be tested with respect to both the Gaussian basis and the plane-wave cutoff. This was done by means of static energy and force calculations on a large-pore configuration of one unit cell of MIL-53(Ga). The configuration was generated by a short $NVT$ run ($T$ = 300 K) starting from the experimental crystal structure[32], with the only aim of creating an out-of-equilibrium structure with non-zero atomic forces. For this configuration, we calculated reference forces and stress tensor components, using very tight computational settings: quadruple-zeta valence triply polarized (QZV3P) Gaussian basis sets[36] for C, H, and O, and the double-zeta valence plus polarization (DZVP) basis set[37], optimized for molecules (MOLOPT), for Ga. A plane-wave cutoff $E_{cut}$ for the density of 2500 Ry was used, as well as a relative cutoff $E_{cut}^{rel}$ of 100 Ry[38], and a convergence criterion for the self-consistent field iterations of $10^{-7}$. The Brillouin zone was sampled at the Γ point only. The interactions between ionic cores and valence electrons were represented by GTH pseudopotentials[39–41], and the exchange and correlation energies were approximated by the PBE functional[42].

The convergence of the atomic forces with respect to the size of the Gaussian basis sets was tested by performing static energy/force calculations with the same settings as in the reference calculation, except for the basis sets of C, O, and H, which were chosen as DZVP, TZVP, or TZV2P, thus increasingly extending their size. For Ga, the DZVP-MOLOPT basis set was used throughout. The calculated atomic forces are plotted against the reference in Fig. 1 (for the sake of clarity, we only show the $x$ components). It can be seen that the DZVP and TZVP basis sets are not able to reproduce the reference results, and only the TZV2P basis sets yield good agreement. The average relative errors for the forces and for the pressure are shown in Fig. 2, and only with the TZV2P basis sets, the error could be reduced to an acceptable level of 1.1% for the forces and to 2.8% for the pressure. Note that the pressure is more difficult to converge than the atomic forces and requires the use of a large Gaussian basis in $NPT$ simulations: even with the TZVP basis sets, it differs from the converged result by more than 40%.

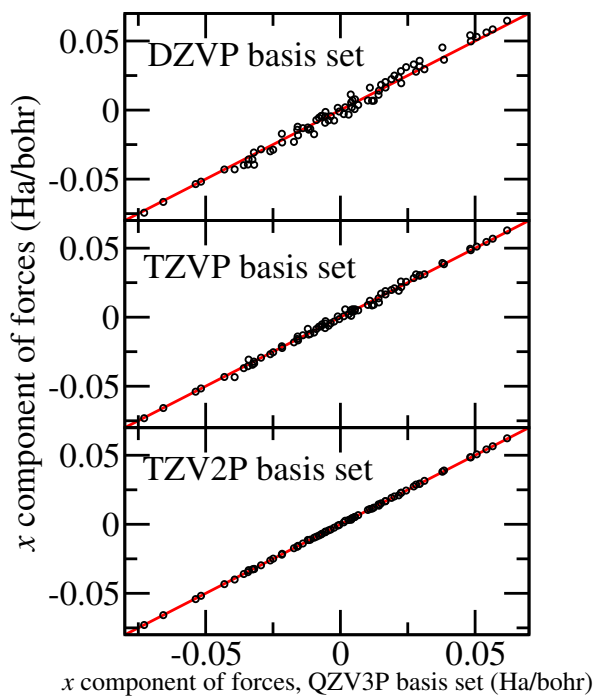

Figure 1. Convergence of forces with basis set size, using a very large cutoff ($E_{cut}$ = 2500 Ry, $E_{cut}^{rel}$ = 100 Ry).

Using the TZV2P basis sets (DZVP-MOLOPT for Ga), we then checked, in a second step, the convergence of forces and pressure as a function of the plane-wave cutoff. Again, static energy/force calculations were performed with the PBE exchange-correlation functional and for the out-of-equilibrium large-pore configuration described above. A TZV2P calculation with $E_{cut}$ = 2500 Ry and $E_{cut}^{rel}$ = 100 Ry served as the reference. In Tab. I, we show the mean realtive error of forces by element using two smaller cutoffs, both with $E_{cut}^{rel}$ = 40 Ry. With $E_{cut}$ = 280 Ry, the default value of the software which is routinely used for production runs, reasonably converged forces are obtained for C, H and O, but the forces acting on Ga are off by almost an order of magnitude. For the pressure, a relative error of 30% is obtained. Only by increasing the cutoff to 600 Ry, acceptable agreement with the reference calculation could be achieved, with a global relative error of 1% and 0.7% for the forces and the pressure, respectively. In summary, TZV2P basis sets (DZVP-MOLOPT for Ga), combined with a plane-wave cutoff of 600 Ry, were found necessary to obtain converged forces and pressures in MIL-53(Ga). These settings were used in the remainder of the present article. Furthermore, we also checked how the restriction to the Γ point affects the stress tensor. Upon doubling the simulation cell and thus improving the sampling of the Brillouin zone, its diagonal elements for the out-of-equilibrium configuration change by less than 10%, and its off-diagonal elements by less than 5%. The use of a single unit cell therefore represents a reasonable trade-off between accuracy and computational efficiency.

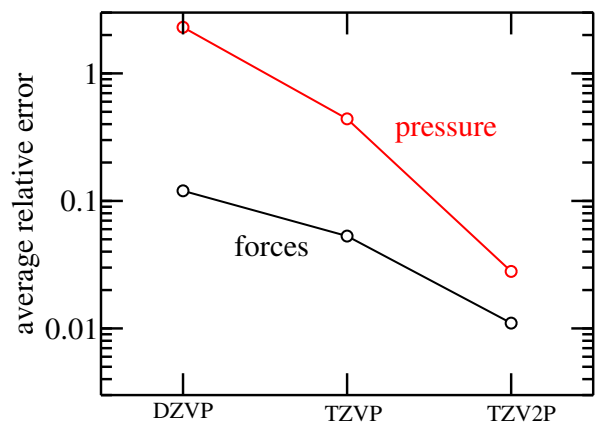

Figure 2. Relative average error of forces and pressure as a function of basis set size, using a very large cutoff ($E_{cut}$ = 2500 Ry, $E_{cut}^{rel}$ = 100 Ry).

|    | 280 Ry | 600 Ry |
|----|--------|--------|
| Ga | 607%   | 10.7%  |
| O  | 12.7%  | 4.2%   |
| C  | 0.64%  | 0.02%  |
| H  | 0.11%  | 0.00%  |

Table I. Convergence of forces (mean relative error) by element with respect to the plane-wave cutoff. As reference, a cutoff of 2500 Ry was used.

In the QUICKSTEP module of CP2K, the computation of Coulomb and exchange-correlation energies is based on a real-space grid, and the density of grid points determines the auxiliary plane-wave cutoff for the representation of the elctronic density. In simulations with variable cell size, one has two options: Either one keeps the number of grid points in the cell constant, which makes the plane-wave cutoff effectively cell size-dependent and biases total energies (and hence *NPT* sampling statistics) for different cell volumes. Or the density of grid points (i.e. the plane-wave cutoff) is kept constant, which can lead to spurious jumps in computed quantities when grid points leave or enter the simulation cell. In a pilote study on pure water, McGrath et al. found that these jumps decrease with increasing plane-wave cutoff and recommend the latter option[43]. Since we used a relatively large cutoff of 600 Ry, we chose to follow this procedure rather than constraining the number of grid points in the simulation cell.

The QUICKSTEP module of CP2K offers the possibility to smooth the electronic density for the calculation of the exchange-correlation energy[36] which may in some cases improve the convergence of total energies and forces with respect to the plane-wave cutoff. The reason behind this option is that the used numerical implementation of DFT breaks the translational invariance of the system. This can lead to spurious forces on atoms, in particular at small cutoffs. However, we noticed that at least in the present case, different tested smoothing schemes lead to changes in total energies of 1.6 eV up to more than 8 eV, and to changes in pressure between 10% and 40%, even at a large plane-wave cutoff of 2500 Ry. Therefore, we did not apply any smoothing and conclude that, while it may be beneficial for the evaluation of forces[44], smoothing is best avoided for *NPT* simulations.

## IV. DISPERSION INTERACTIONS

### A. Grimme correction with original parameters

Local or semi-local exchange correlation functionals like PBE are known to poorly represent dispersion interactions. Since these are expected[12] (and shown below) to play a critical role in MIL-53(Ga), we tested two approaches to take them into account: the first one, proposed by Grimme[45], is based on a pair potential which is added on top of the local or semi-local DFT scheme. The second, due to Dion *et al.*[46], incorporates dispersion interactions directly in the DFT framework by using a non-local exchange-correlation functional. To assess the validity of the different approaches, we performed MD simulations in the *NPT* ensemble at different temperatures and compared the resulting structures to those observed experimentally.

We first tested the Grimme dispersion correction in its original form ("Grimme D2")[45], which adds a parameterized attractive interaction $\propto 1/R^6$, damped at short distances, to the DFT total energy (with $R$ denoting the distance between two atoms). This interaction is scaled by a global factor $s_6$ which only depends on the used exchange-correlation functional and takes the value 0.75 for the PBE functional. With these settings, we performed an *NPT* simulation with a temperature of 600 K and a pressure of 1 bar, starting from a large-pore configuration of the material. In these conditions, the large-pore form was observed to be stable in experiments[33]. In the simulation, temperature was controlled by a Nosé-Hoover thermostat[47,48] with a time constant of 100 fs, and a barostat[49] with a time constant of 2 ps allowed for changes in size and shape of the simulation cell. We checked the possible influence of the simulation protocol on the results by performing simulations with various time constants for the barostat and the thermostat and also with a different thermostat[50]. These choices were found to affect volume and temperature fluctuations as well as the time scale of structural transformations, but the conlusions regarding the relative stability of the different phases turned out to be independent of the details of the barostat and thermostat settings.

Fig. 3 (solid lines) shows the evolution of the lattice parameters *a*, *b* and *c* during this simulation. The parameter *b* decreases to approximately half its original value, while *a* increases slightly and *c*, pointing along the inorganic –OH–Ga–OH– chains, does not change significantly. This evolution reflects a transition to a structure with virtually closed pores ($b < c$, see Fig. 4), even narrower than seen experimentally in the narrow-pore phase at room temperature ($b > c$). We therefore conclude that the Grimme correction in its original form, combined with the PBE exchange-correlation functional, does not describe the interactions in MIL-53(Ga) properly. In particular, it overestimates the dispersion interactions between the organic linkers, forcing them into a very compact configuration. While such a "very narrow pore" form has been observed in the scandium-bearing MIL-53(Sc)[51], it does not occur for MIL-53(Ga). For completeness, we also tested the more recent version of the Grimme dispersion correction ("Grimme D3")[52], which remedies the trend for overbinding of the original one in many cases. However, we still obtained a collapse of the large-pore phase, albeit somewhat slower than in the first case (Fig. 3, dashed lines). Hence, the Grimme dispersion correction in its two flavors is not suited for simulating the high-temperature large-pore form, nor does it give the correct structure of the narrow-pore phase.

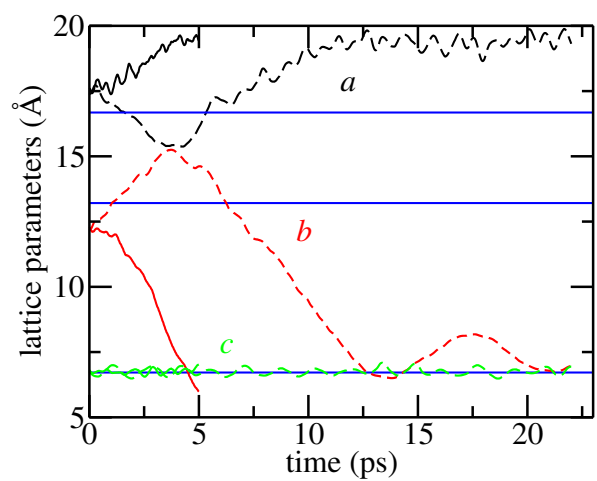

Figure 3. Evolution of the cell parameters at 1 bar, 600 K, with the Grimme dispersion correction (solid line: D2[45], dashed line: D3[52]). The experimental lattice parameters[33] for the large-pore phase are shown as blue lines.

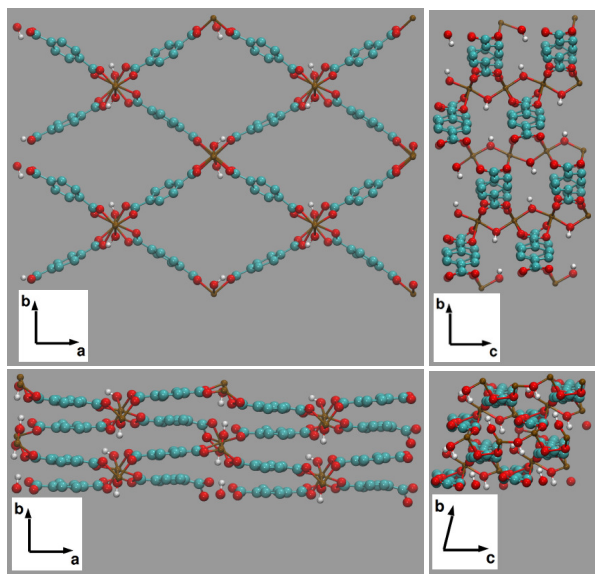

Figure 4. Snapshots from the beginning (upper panel) and the end (lower panel) of the *NPT* simulation (*P* = 1 bar, *T* = 600 K) with the original Grimme D2 dispersion correction and PBE. The simulation box consists of one unit cell and is repeated periodically here by 2×2×2 units for a better visualization of the structure. For clarity, hydrogen atoms bonded to the benzenedicarboxylate linkers are not shown.

### B.  Varying the global scaling parameter of the Grimme correction

The global scaling factor of the Grimme D2 dispersion correction, $s_6$ = 0.75 for PBE, was originally determined by optimizing binding energies of a set of 40 noncovalently bound complexes and can thus be considered an adjustable parameter, representing a compromise for a large range of systems and configurations[45]. Since the dispersion correction was found to overestimate the attractive interactions between the organic linkers in MIL-53(Ga), we decreased the interaction strength, i.e. $s_6$, in small steps. At each step, we performed *NPT* simulations at 1 bar and 600 K, starting from a large-pore configuration and using the same settings as earlier.

In all simulations with $s_6$ > 0.2, the initial structure transformed to a narrow-pore phase, similarly to what was observed in Section IV A. However, with $s_6$ = 0.2, the large-pore structure was stable throughout a 15 ps simulation. Taking the first 5 ps as an equilibration run, we plot the evolution of the cell parameters over the final 10 ps in Fig. 5. The average lattice parameters and cell angles are in good agreement with the experimental ones (see Table II), especially if one bears in mind the softness of the material[53] which can be linked to the large fluctuations of the lattice parameters over time (Fig. 5). We conclude from this that the Grimme dispersion correction with $s_6$ = 0.2 and the PBE functional describes the large-pore phase correctly.

However, when using the same simulation settings, the narrow-pore phase, which is experimentally observed to be stable between 350 K and 500 K[33], is not modelled correctly. Fig. 6 shows the lattice parameters of MIL-53(Ga) during an

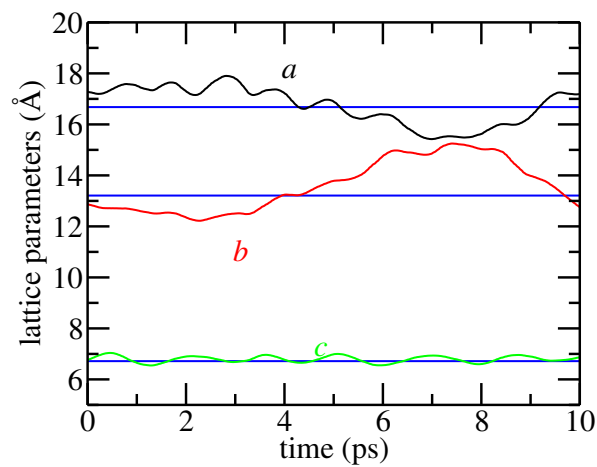

Figure 5. Evolution of the cell parameters at 1 bar, 600 K, with the modified Grimme dispersion correction (global scaling factor $s_6$ = 0.2). The experimental lattice parameters[33] for the large-pore phase are shown as blue lines.

*NPT* simulation at 1 bar and 373 K, starting from a narrow-pore configuration. The material clearly undergoes a transition to a large-pore structure with subsequent large-amplitude fluctuations, in contradiction with experiment. It is concluded that this modified Grimme dispersion correction is not adequate for simulating the narrow-pore phase.

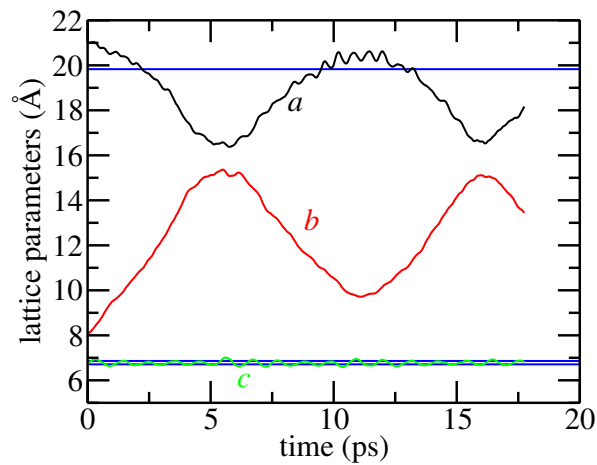

Figure 6. Evolution of the cell parameters at 1 bar, 300 K, with the modified Grimme dispersion correction (global scaling factor $s_6$ = 0.2). The experimental lattice parameters[33] for the narrow-pore phase are shown as blue lines.

### C.  Non-local exchange-correlation functional including dispersion

So far, we have shown that the large-pore phase of MIL-53(Ga) can be modelled with a modified Grimme dispersion correction. On the other hand, the narrow-pore form is not correctly described by the original Grimme dispersion correc-tion, which yields an over-compact structure, nor by modified schemes fine-tuning the global scaling factor to $s_6$ = 0.2, which lead to the opening of the pores. We thus explored a second promising approach to incorporate dispersion interactions in the framework of DFT, which consists in the use of a non-local exchange-correlation functional due to Dion *et al.*[46], written as

$$E_{xc} = E_x^{revPBE} + E_c^{LDA} + E_c^{non-local} \quad (1)$$

|  | a (Å) | b (Å) | c (Å) | α (deg) | β (deg) | γ (deg) | volume (Å³) |
|---|---|---|---|---|---|---|---|
| Large-pore | | | | | | | |
| sim. | 16.73 | 13.59 | 6.79 | 91 | 90 | 90 | 1535.8 |
| exp.[33] | 16.68 | 13.21 | 6.72 | 90 | 90 | 90 | 1479.7 |
| Narrow-pore | | | | | | | |
| sim. | 19.49 | 6.99 | 6.83 | 90 | 98 | 90 | 919.49 |
| exp.[33] | 19.83 | 6.86 | 6.71 | 90 | 104 | 90 | 886.28 |

Table II. Structural parameters of the empty large- and narrow-pore phase of MIL-53(Ga), obtained from first-principles molecular dynamics simulations in the $NPT$ ensemble. Large-pore simulation: 600 K, 1 bar, modified Grimme dispersion correction ($s_6$ = 0.2). Narrow-pore simulation: 300 K, 1 bar, non-local Dion functional.

It combines, on the right-hand side of Equation 1, the exchange part of the revPBE functional[54], the correlation energy functional in the local density approximation[55] and an additional non-local correlation energy functional which takes into account dispersion interactions. With our hardware and software setup, we found this functional to be computationally more expensive by a factor of 1.5 compared to the local DFT-D2/3 functional.

We used this non-local exchange-correlation functional for an $NPT$ simulation at 300 K and 1 bar, starting from a narrow-pore configuration. Temperature was controlled by the thermostat proposed by Bussi *et al.*[50], with a time constant of 100 fs, and the barostat time constant was 2 ps. It can be seen from Fig. 7 that with these settings, the narrow-pore phase is stable during the simulation (as it should), and the average cell parameters agree well with the experimental ones (see Table II), although the angle $\beta$ of the monoclinic unit cell is only 98°, instead of the measured 104°.

However, the large-pore form is not correctly described with this functional, at least in its present form (Equation 1), as shown in Fig. 8 (full lines): at 600 K and 1 bar, where it is observed to be stable in the experiment, it undergoes a transition to the narrow-pore form in the $NPT$ simulation. Now, it could be conjectured that the erroneous transition in the simulation may be due to the small size of the simulation box which consists of one unit cell of MIL-53(Ga). The finite size may affect the location of the phase transition since it limits the available vibrational modes and hence might have an impact on the free energy (in principle, finite size can influence also the static stress tensor via the sampling of the Brillouin zone, but we have shown in section III that the used Γ point sampling has only a small effect). We therefore performed the same simulation using a box of twice the original size, doubling the lattice parameter *c*. Also the larger simulation box leads to the (unphysical) closure of the pores, although the structural transition is slowed down with respect to the simulation of a single unit cell (Fig. 8, dashed lines). This shows that the finite size influences the kinetics of the process, but not its (unphysical) equilibrium state.

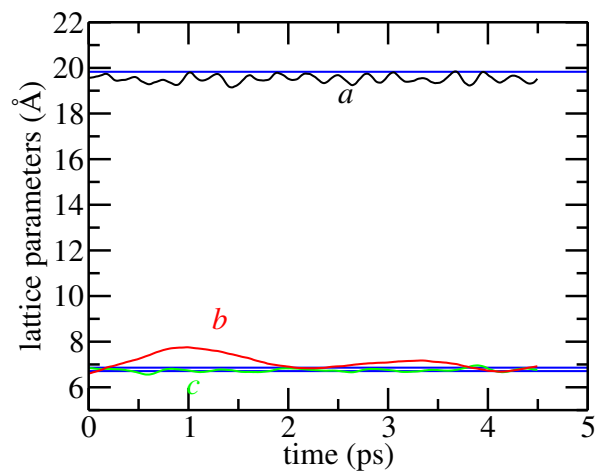

Figure 7. Evolution of the cell parameters of the narrow-pore phase, at 1 bar and 300 K, with the Dion exchange-correlation functional. The experimental lattice parameters[33] for the narrow-pore phase are shown as blue lines.

## V. SIMULATION OF THE HYDRATED MIL-53(GA)

Materials of the MIL-53 family exhibit pores large enough to accomodate small guest molecules, and it is their role as adsorbants which makes them interesting from the point of view of practical applications. Given the result that the non-local exchange-correlation functional yields the correct structure of the empty narrow-pore phase, we checked if the same scheme could also describe the hydrated narrow-pore form correctly. It is experimentally stable under ambient air up to 350 K (where dehydration starts), contains one water molecule per Ga and exhibits slightly wider pores than the empty material[33].

The $NPT$ simulations were performed at 300 K and at 1 bar, using a Bussi thermostat and a barostat with time constants of 100 fs and 2 ps, respectively. We started from a narrow-pore form into which four water molecules were inserted close to the $\mu_2$-OH groups of the framework, such that hydrogen bonds were obtained between the water oxygen and $\mu_2$-OH, as observed in x-ray diffraction[33]. After 3.5 ps of equilibration, the average cell parameters were obtained from 7 ps MD and are listed in Table III. Reasonable agreement with experiment was found, and the largest difference was observed for the lattice parameter *b*, which is overestimated by the simulation by 0.5 Å. However, one should bear in mind that *b* corresponds to the soft direction that governs the opening of the pores and is therefore extremly sensitive to temperature, pressure and the theoretical model.

|      | a (Å) | b (Å) | c Å  | α (deg) | β (deg) | γ (deg) | volume (Å³) |
|------|-------|-------|------|---------|---------|---------|-------------|
| sim. | 19.38 | 8.10  | 6.79 | 90      | 97      | 90      | 1056        |
| exp.[33] | 19.72 | 7.58 | 6.69 | 90   | 103     | 90      | 972         |

Table III. Structural parameters of the hydrated narrow-pore phase of MIL-53(Ga), obtained from first-principles molecular dynamics simulations in the *NPT* ensemble.

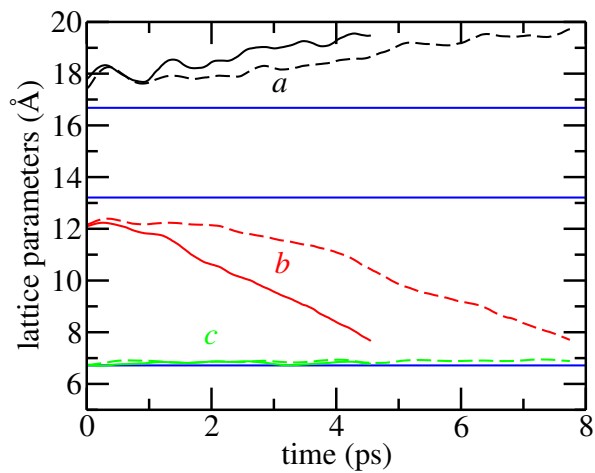

Figure 8. Evolution of the cell parameters at 1 bar, 600 K, with the Dion exchange-correlation functional. Full and dashed lines represent simulations using a single unit cell and a $1 \times 1 \times 2$ supercell, respectively. The experimental lattice parameters[33] for the large-pore phase are shown as blue lines.

## VI. CONCLUSIONS

In conlusion, we have shown that MIL-53(Ga) is a challenging system for molecular dynamics with flexible size and shape of the simultion cell. We presented simulation schemes which successfully reproduce experimental results. The calculation of accurate stresses and atomic forces, especially on Ga, require the use of large Gaussian basis sets and plane-wave cutoffs, and widely used standard settings are not sufficient. Moreover, the softness of the material makes its equilibrium structure and the location of phase transition very sensitive to the density functional chosen in the simulation. In particular, dispersion interactions were shown to play a crucial role in determining the transition between the narrow- and the large-pore phase. In order to correctly describe the material's behavior, the theoretical model has to capture the delicate balance between dispersion forces, preferring a narrow-pore form, and coordination chemistry of Ga, which favors the large-pore structure.

In view of these intricacies, no single simulation scheme could be identified which would allow a unified description of both the large- and narrow-pore phase of MIL-53(Ga). Instead, we propose to use the PBE functional in conjunction with a modified Grimme dispersion correction with a global scaling parameter $s_6 = 0.2$ for the high-temperature, large-pore form. These settings can be used, e.g., for studying the hydrothermal stability of the material. On the other hand, the non-local exchange-correlction functional proposed by Dion *et al.* seems to be very promising: it gives the correct structures of the narrow-pore form, both empty and hydrated, and is thus suited for simulating the low-temperature adsorption of guest molecules and its interplay with the structural flexibility of MIL-53(Ga). Moreover, we expect the Dion functional to be useful also for the strongly hydrated large-pore form of MIL-53 materials in which the water molecules filling the pores prevent the structure from collapsing to the narrow-pore phase[56].

## VII. ACKNOWLEDGEMENTS

The authors would like to thank Aurélie Ortiz for valuable discussions about the Grimme dispersion correction scheme. This work was supported by the Agence Nationale de la Recherche under project ''SOFT-CRYSTAB'' (ANR-2010-BLAN-0822), and performed using high-performance computing resources from GENCI-IDRIS (grants i2013086882 and x2013086114).